\documentclass[preprint,nofootinbib,aps,superscriptaddress,eqsecnum]{revtex4-1} 
\pdfoutput=1
\usepackage{extarrows}
\usepackage{graphicx}
\usepackage{amsmath}
\usepackage{amsfonts}
\usepackage{amssymb}
\usepackage{hyperref}
\usepackage{float}
\usepackage[utf8]{inputenc}
\usepackage{caption}
\usepackage{subcaption}
\usepackage{color}

\allowdisplaybreaks
\def\bea{\begin{eqnarray}}
	\def\eea{\end{eqnarray}}
\def\be{\begin{equation}}
	\def\ee{\end{equation}}

\begin{document}

	\title{Neutrinoless double beta decay and Sterile dark matter in extended left right symmetric model}
	\author{Bichitra Bijay Boruah}
	\email{bijay@tezu.ernet.in}
	\author{Nayana Gautam}
	\email{nayana@tezu.ernet.in}
	\author{Mrinal Kumar Das}
	\email{mkdas@tezu.ernet.in}
	\affiliation{Department of Physics, Tezpur University, Tezpur 784028, India}
	\begin{abstract}
		
		We have studied a flavor symmetry based extended left-right symmetric model(LRSM) with dominant type-II seesaw
		mechanism and have explored the associated neutrino phenomenology. The particle content of the model includes usual quarks, leptons along with additional sterile fermion per generation in the fermion sector while the scalar content contains
		Higgs doublets and scalar bidoublet. Realization of this extension of LRSM has been done by using $A_{4}\times Z_{4}$ discrete symmetries. In this work, we have also included the study of sterile neutrino dark matter(DM) phenomenology along with neutrinoless double beta decay within the framework.
	\end{abstract} 
	\maketitle

		\section{INTRODUCTION}
			Standard Model(SM) is the most economical and successful model of particle physics, But it fails to explain various phenomena like lepton flavor violation, small neutrino mass, baryon asymmetry of universe, dark matter etc. One of the main drawback of the SM is that it explains the parity violation of weak interaction adhocly by assuming that transformation of right handed counter part of the particles to be transformed as singlet under $SU_{2}$ gauge group. Because of this, there has to be some framework where parity violation can be explained in a natural way. Left right symmetric model is one such framework where both right handed and left handed particle are treated on equal footing and it is the most economical way to restore parity came a long way in explaining neutrino mass, lepton number violation, dark matter, baryon asymmetry of the universe thereby gaining popularity. LRSM \cite{Mohapatra:2006xy,Mohapatra:1986aw,Mohapatra:1974gc,Pati:1974yy,Senjanovic:1975rk}  is a very simple extension of the standard model gauge group where parity restoration is obtained at a high energy scale and the fermions are assigned to the gauge group  $ SU(3)_c\times SU(2)_L\times SU(2)_R\times U(1)_{B-L}$ which can be tested in present-day experiments. The usual type-I\cite{Minkowski:1977sc,Schechter:1980gr,Mohapatra:1979ia,Yanagida:1979as,Gell-Mann:1979vob} and type-II\cite{Cheng:1980qt,Magg:1980ut,Lazarides:1980nt,Mohapatra:1980yp,Wetterich:1981bx,Antusch:2004xy} seesaw arises naturally in LRSM. Several other problems like parity violation of weak interaction, massless neutrinos, CP problems, hierarchy problems, etc can also be addressed in the framework of LRSM. The seesaw scale is identified as the breaking of the $SU(2)_R$ symmetry. In this model, the electric charge generator is given by, $ Q=T_{3L}+T_{3R}+\frac{B-L}{2}$, where $ T_{3L}$ and $ T_{3R}$ are the generators of $ SU(2)_L$ and $ SU(2)_R$ and B-L being the baryon minus lepton number charge operator. Mass of neutrino can be explained in LRSM with canonical seesaw, inverse seesaw, linear seesaw, extended seesaw etc. In this work LRSM  is extended with a neutral fermions per generation, which is known as extended seesaw. Within this extended seesaw type-II dominance is considered for phenomenological analysis. With this extra singlet fermion the mass matrix in type-II dominance can be given as-
		\begin{equation}
		\mathcal{M} = \begin{pmatrix}
		M_{L} & 0 & M_{D} \\
		0 & 0 & M\\
		M_{D}^{T} & M^{T} & M_{R}\\
		\end{pmatrix}\label{eq21}
		\end{equation}
		Where light neutrino mass is $m_{\nu}=\nu_{L}f_{L}$. A natural type-II seesaw  dominance allows large light-heavy neutrino mixing thus leading
		to many new physics contributions to neutrinoless double beta decay along with constraints on light neutrino mass. In one of the recent work \cite{Sahu:2022xkq} extended LRSM is studied in details considering linear seesaw to show the phenomenological importance in different LFV processes and unitarity violation etc using flavor symmetry. 
		
		The nature of the neutrinos and absolute neutrino mass scale is still unknown, whether they are four-component Dirac particles possessing a conserved lepton number or two-component Majorana particles. This is directly related to the issue of LN conservation. One such process of fundamental importance in particle physics which arises in any BSM framework is neutrinoless double beta decay(NDBD/0$\nu$$\beta$$\beta$). It is a second-order, and slow radioactive process that transforms a nuclide
		of atomic number Z into its isobar with atomic number Z+2,
		\begin{center}
			\begin{equation}
			N(A,Z) \longrightarrow N(A,Z+2)+e^{-}+e^{-}
			\end{equation}
		\end{center}
		which violates the lepton number(LN) conservation. The main aim for the search of 0$\nu$$\beta$$\beta$ decay is the measurement of the effective Majorana neutrino mass, which is a combination of the neutrino mass eigenstates and neutrino mixing matrix terms.
		
		 There is no convincing experimental evidence of the decay that exists
		to date. But the new generation of experiments are already running or about to run to explore effective neutrino mass along with decay rates NDBD process. In addition, from the lifetime of NDBD combined with sufficient knowledge of the nuclear matrix elements (NME), one can set a constrain involving the neutrino masses. The experiments that have improved the lower bound of the half-life of the decay process include KamLANDZen\cite{KamLAND-Zen:2016pfg} and GERDA\cite{gerda} which use Xenon-136 and Germanium-76 nuclei respectively. Incorporating the results from the first and second phase of the experiment, KamLAND-Zen imposes the best lower limit on the decay half-life using Xe-136 as $T_{1/2}^{0\nu}$$>$ $1.07 \times 10^{26}$yr at $90$ percent CL and the corresponding upper limit of effective Majorana mass in the range $(0.061-0.165)$eV. NDBD is extensively studied within the framework of LRSM \cite{Mohapatra:1981pm,Picciotto:1982qe,Hirsch:1996qw,SuperNEMO:2010wnd,Tello:2010am,Chakrabortty:2012mh,Nemevsek:2012iq,Patra:2012ur,Barry:2013xxa,BhupalDev:2013ntw,BhupalDev:2014qbx}. New contribution to NDBD process in extended seesaw mechanism also studied considering different type of dominance such as inverse, linear etc in many literatures\cite{Sruthilaya:2017vui,Senapati:2020alx,Deppisch:2014zta,Parida:2018apw,Awasthi:2013ff}. In this work we aim to discuss the extended contrbution to NDBD process using flavor symmetry.
		
		The dark matter(DM) and its nature are the most important and open questions in particle physics community till date. Different astrophysical and cosmological measurements confirm that some part of our present Universe is made up of DM, which are basically non-luminous and non-baryonic. Cosmic microwave background(CMB)power spectrum, cluster and galactic rotation curves and gravitational lensing etc are some astrophysical measurements that confirms the existence of the DM. Although the presence of the DM is confirmed from above mentioned measurement, the nature and components of DM is still unknown. Almost  26.8 \% of the energy density of the universe is made up of DM according to Planck data and the dark matter abundance is \cite{ade2016ade} :
		\begin{equation*}
		\Omega_{DM}h^2 = 0.1187 \pm 0.0017 
		\end{equation*}
		
		Connecting BSM physics to DM with other unsolved problems like the smallness of neutrino mass, BAU, etc has been a  great challenge to the physics community \cite{mass-mixingneutrino,king2014neutrino}. From the observational evidences till date, it is clear that DM candidates must be massive, neutral to the electromagnetic interactions, and stable. However,
		Standard Model fails to provide a viable DM candidate. The particle physics community has begun to study different possible BSM frameworks which can give rise to the correct DM phenomenology. A plethora of DM candidates has been proposed within different BSM frameworks. As proposed in several works, sterile neutrinos in keV scale can be a well-motivated feasible dark matter candidate \cite{adhikari2017white,merle2017kev,lucente2016implication,admixturesterile,kusenko2009sterile,abazajian2017sterile,adhikari2017white,merle2017kev,lucente2016implication,admixturesterile,kusenko2009sterile,abazajian2017sterile,upperlimit,mixinglimit}. Sterile neutrinos are right-handed gauge singlet neutrinos that posses gravitational interactions only. Even in the X-ray band, one of the most studied candidates of dark matter is sterile neutrino, which can radiatively decay into an active neutrino and a mono energetic photon in the process $N\longrightarrow\nu+\gamma$.
		
		In this work, we have used $A_{4}\times Z_{4}$ discrete symmetries to realize the extended seesaw with type-II dominance. We aim to study here the extension of LRSM which offers the advantage of
		studying neutrino mass, new physics contribution to NDBD process and DM phenomenology  within one framework. $A_{4}$ is the discrete group of even permutations of four objects
		is the smallest non-Abelian group containing triplet irreducible representations. More importantly, flavor symmetry has been used within the framework of LRSM in very few works \cite{Rodejohann:2015hka,Bonilla:2020hct,Bazzocchi:2007au,Boruah:2021ktk}. In some other important literatures extended LRSM is also dicussed considering different dominant seesaw\cite{Sruthilaya:2017vui,Parida:2016hln,Parida:2018apw,Sahoo:2017cqg,Chakraborty:2019uxk}.  We extensively discussed different leading order contributions to NDBD process and DM phenomenology in the considered framework. Discrete symmetry will constrain different mass matrices arising from framework considered and which will lead to phenomenological implications. 
		
	     This paper is structured as follows. In section \ref{sec2}, we briefly discuss the left-right symmetric model with extended seesaw with type-II dominance. Flavor symmetric model considering type-II dominance are discussed in section \ref{sec3}. We also discuss the different new physics contribution to the amplitude of the decay process in section \ref{sec4}.
		In section \ref{sec5}, we discuss light sterile neutrino as a probable dark matter candidate. We present our numerical analysis and results in section \ref{sec6} and then in Section \ref{sec7},
		we conclude by giving a brief overview of our work.

		\section{LRSM with extended seesaw with type-II dominance:\label{sec2}}
		In our work LRSM is extended with a neutral sterile fermion per generation along with quark and leptons. Scalar sector of this extension consist of Higgs bidoublet $\Phi$, two scalar triplets $\Delta_{L}$, $\Delta_{R}$ and two doublets $H_{L}$,$H_{R}$. The B-L charge of Higgs bidoublet, scalar triplets and doublets are  0,2 and -1 respectively. This kind of extension of LRSM is known as extended LR model and naturally arising seesaw from it is called the extended seesaw mechanism which is well explained in literaures such as \cite{Hati:2018tge,Sruthilaya:2017vui} . It is quite well known that type-II dominance in such scenario allows large left right mixing which leads to interesting phenomenology. 
		The Quarks(Q) and leptons(l) (LH and RH) that transform in L-R symmetric gauge group are given by,
		
		\begin{equation}\label{eqx}
			Q_L=\left[\begin{array}{cc}
				u^{'}\\
				d^{'}
			\end{array}\right]_L,Q_R=\left[\begin{array}{cc}
				u^{'}\\
				d^{'}
			\end{array}\right]_R, l_L=\left[\begin{array}{cc}
				\nu_l\\
				l
			\end{array}\right]_L,l_R=\left[\begin{array}{cc}
				\nu_l\\
				l
			\end{array}\right]_R.
		\end{equation}
	
	where the Quarks are assigned with quantum numbers $(3,2,1,1/3)$ and $(3,1,2,1/3)$ and leptons with $(1,2,1,-1)$ and $(1,1,2,-1)$ respectively under
	$ \rm SU(3)_c\times SU(2)_L\times SU(2)_R\times U(1)_{B-L}$. The sterile fermion transformation under the relevant Gauge group is given as $S_{L}(1,1,1,0)$ The scalar sector in this extension of LRSM consists of a bi-doublet with quantum number $ \rm \Phi(1,2,2,0)$
	, the $SU(2)_{L,R}$ triplets, $\Delta_L(1,3,1,2)$, $\Delta_R(1,1,3,2)$ and two $SU(2)_{L,R}$ doublets, $H_L(1,2,1,-1)$, $H_R(1,1,2,-1)$ . 
	
	A natural type-II dominance in such case can be realised with the following Yukawa interaction-
	
	\begin{align}
		 -\mathcal{L_{Y}}=\Bar{l_{L}}[Y_{1}\Phi+Y_{2}\tilde{\Phi}] l_{R}+
		f[\Bar{l_{L}}^{c} l_{L}\Delta_{L}+\Bar{l_{R}}^{c} l_{R}\Delta_{R}]+F(\Bar{l_{R}})H_{R}S_{L}^{c}+h.c \label{eq1} 
	\end{align}
	
In terms of mass matrices we can write Eq\eqref{eq1} as-
\begin{align}
	-\mathcal{L_{Y}} \supset M_{D} \Bar{\nu_{L}} N_{R}+M_{L}\Bar{\nu_{L}}^{c}\nu_{L}+M_{R}\Bar{\nu_{R}}^{c}\nu_{R}+M\bar{N_{R} S_{L}}+h.c 
\end{align}
where, $M_{D}=Y<\Phi>$ is the Dirac neutrino mass matrix which measures the light and heavy neutrino mixing and Y stands for Yukawa coupling, $M_{R}=f<\Delta_{R}>=f\nu_{R}$ ($M_{L}=f<\Delta_{L}>=f\nu_{L}$) is the Majorana mass term for heavy(Light) neutrinos and $f$ is the Majorana left right symmetric coupling along with $\nu_{R}$, $\nu_{L}$ corresponding to the vacuum expectation values(VEV) of scalar triplets, finally $M=F<H_{R}>$ stands for heavy neutrino-sterile mixing matrix. We have assumed the induced vev of $<H_{L}>$ to be zero. Now in flavor basis the complete $9\times 9$ mass matrix for neutral fermions can be represented as-

\begin{equation}
	\mathcal{M} = \begin{pmatrix}
		M_{L} & 0 & M_{D} \\
		0 & 0 & M\\
		M_{D}^{T} & M^{T} & M_{R}\\
	\end{pmatrix}\label{eq2}
\end{equation}

One can use standard seesaw mechanism to integrate out the heaviest neutrino, Using Mass ordering $M_{R}>M>M_{D}\gg M_{L}$ one can obtain-
\begin{equation}
	\mathcal{M}^{'} = \begin{pmatrix}
		M_{L}-M_{D}M_{R}^{-1}M_{D}^{T} &- M_{D}M_{R}^{-1}M_{D}^{T}  \\
		M M_{R}^{-1}M_{D}^{T} & M M_{R}^{-1}M^{T}\\
	\end{pmatrix}\label{eq3}
\end{equation}

In the Eq\eqref{eq3} which is the intermediate block diagonalised mass matrix, it is seen that (2,2) element is larger than the entries in the limit  $M_{R}>M>M_{D}\gg M_{L}$. So using the same procedure we can integrate out sterile neutrino mass matrix. Now we can write down light neutrino mass matrix as-
\begin{align}
	\begin{split}
	 & m_{\nu} =[M_{L}-M_{D}M_{R}^{-1}M_{D}^{T}]-(-M_{D}M_{R}^{-1}M_{D}^{T})(M M_{R}^{-1}M_{D}^{T}) (M M_{R}^{-1}M^{T})\\ 
	 &=[M_{L}-M_{D}M_{R}^{-1}M_{D}^{T}]-M_{D}M_{R}^{-1}M_{D}^{T}\\
	 &=M_{L}=m_{\nu}^{II}
   \end{split}\label{eq4}
\end{align}

After complete block diagonalisation which is extensively discussed in various literatures, we can write t left handed neutrino, heavy right handed neutrino and sterile neutrino mass matrices as-
\begin{align}
	\begin{split}
	& m_{\nu}= M_{L}\\
	&M_{R}= \frac{\nu_{R}}{\nu_{L}}M_{L}\\
    &M_{S}=- M M_{R}^{-1}M^{T}
	\end{split}\label{eq5}
	\end{align}
Again these mass matrices can be diagonalised by using respective $3\times3$ unitary matrices as follows-
\begin{align}
	\begin{split}
		& m_{\nu}^{diag}= U_{\nu}^{\dagger}m_{\nu}U_{\nu}^{*}=diag(m_{1},m_{2},m_{3})\\
		&M_{N}^{diag}= U_{N}^{\dagger}M_{N}U_{N}^{*}=diag(M_{N_{1}},M_{N_{2}},M_{N_{3}})\\
		&M_{S}^{diag}= U_{S}^{\dagger}M_{S}U_{S}^{*}=diag(M_{S_{1}},M_{S_{2}},M_{S_{3}})
	\end{split}
\end{align}

Complete block diagonalization results-
\begin{align}
	\begin{split}
		& \mathcal{M}_{diag}=V _{9\times9}^{\dagger}\mathcal{M}V _{9\times9}^{*}\\
		&=(\mathcal{W}.\mathcal{U})^{\dagger}\mathcal{M}(\mathcal{W}.\mathcal{U})\\
		& =diag(m_{1}.m_{2},m_{3,}M_{S_{1}},M_{S_{2}},M_{S_{3}},M_{N_{1}},M_{N_{2}},M_{N_{3}})
	\end{split}
\end{align}
where, $\mathcal{W}$ is the bock diagonalized mixing matrix and $\mathcal{U}$ is the unitary mixing matrix. Thus the complete $9\times9$ unitary mixing matrix can be written as-
\begin{equation}
	V=\mathcal{W}.\mathcal{U}= \begin{pmatrix}
		U_{\nu} &  M_{D}M^{-1}U_{S} & M_{D}M_{R}^{-1}U_{N} \\
	(M_{D}M^{-1})^{\dagger}U_{\nu}& U_{S} & M M_{R}^{-1}U_{N}\\
		\mathcal{O} & (M M_{R}^{-1})^{\dagger}U_{S} & U_{N}\\
	\end{pmatrix}\label{eq6}
\end{equation}

\section{Flavor symmetric realization of extended LRSM with type-II dominance\label{sec3}:} 

In this section we have realized the extended LRSM considering type-II dominance case using $A_{4}\times Z_{4}$ flavor symmetry. Type-II dominance in extended LRSM allows for large left right mixing. $A_4$ is a discrete group of even permutations of four objects. It has three inequivalent one dimensional representations 1, $1^{\prime}$ and$1^{\prime\prime}$ an irreducible three dimensional representation 3. Product of the singlets and triplets are given by-

\begin{equation}
	1 \otimes 1=1 \nonumber
\end{equation} 
\begin{equation}
	1^{\prime}\otimes 1^{\prime}=1^{\prime\prime} \nonumber
\end{equation}
\begin{equation}
	1^{\prime}\otimes1^{\prime\prime}=1 \nonumber
\end{equation}
\begin{equation}
	1^{\prime\prime} \otimes 1^{\prime\prime}=1^{\prime} \nonumber
\end{equation}
\begin{equation}
	3\otimes3=1\oplus1^{\prime}\oplus1^{\prime\prime}\oplus 3_{A}\oplus 3_{S}
\end{equation}

where subscripts A and S stands for “asymmetric” and “symmetric” respectively. If we
have two triplets ($a_1$, $a_2$, $a_3$) and ($b_1$, $b_2$, $b_3$), their products are given by

\begin{equation}
	1 \approx a_1b_1 + a_2b_3 + a_3b_2 \nonumber
\end{equation}
\begin{equation}
	1^\prime \approx a_3b_3 + a_1b_2 + a_2b_1 \nonumber
\end{equation}
\begin{equation}
	1^{\prime\prime} \approx a_2b_2 + a_3b_1 + a_1b_3 \nonumber
\end{equation}
\begin{equation}
	3_S \approx \left(\begin{array}{c}
		2a_{1}b_{1}-a_{2}b_{3}-a_3b_2\\
		2a_{3}b_{3}-a_{1}b_{2}-a_2b_1\\
		2a_{2}b_{2}-a_{1}b_{3}-a_3b_1 \end{array}\right)\nonumber
\end{equation}
\begin{equation}
	3_A \approx \left(\begin{array}{c}
		a_{2}b_{3}-a_{3}b_{2}\\
		a_{1}b_{2}-a_2b_1\\
		a_{3}b_{1}-a_{1}b_3 \end{array}\right)\nonumber
\end{equation}

The particle content and their respective charge assignments of the model are given in Table\eqref{tab2}.
One can get diagonal charged lepton mass matrix and Dirac neutrino mass matrix by considering a suitable flavon with proper charge assignments. 
Now, as we are interested in type-II dominant case within extended LRSM, we have already discussed in the section\eqref{sec2} that mass matrices of light neutrino, heavy neutrino and sterile neutrino are independent of Dirac neutrino mass matrix (not mixing) allow us to assume flavor independent structure of Dirac neutrino mass matrix. This is well motivated for the reason that we can assume that this kind of extension originates from $SO(10)$ GUT (grand unified theory) with proper choice of intermediate symmetry such as Pat-Salaam symmetry. In this scenario, the charged lepton mass matrix and Dirac neutrino mass matrix can be treated in equal footing and taken to be equivalent to up type quark mass matrix. Main objective of this work to constrain the mass matrix of light neutrino, heavy neutrino and sterile neutrino with $A_{4}\times Z_{4}$ flavor symmetry study the impact of this constrained mass matrices on mixing where Dirac neutrino mass matrix plays a crucial role. The light neutrino mass formula is governed by type-II seesaw mechanism, $m^{II}_{\nu}=M_{L}$ while
type-I seesaw contribution $m^{I}_{\nu}=-M_{D}M^{-1}_{R}M_{D}^{T}$ is exactly canceled out in the complete
diagonalization method as given in \cite{Sruthilaya:2017vui}. Since light neutrino mass formula is independent of Dirac neutrino
mass matrix, any value of $M_{D}$ is allowed consistent with GUT mass fitting without any fine
tuning of the Yukawa couplings. Further we study their impact on neutrinoless double beta decay contributions and active sterile mixing with dark matter phenomenology.

\begin{table}[H]
	\begin{center}
		\begin{tabular}{|c|c|c|c|c|c|c|c|c|c|c|c|c|c|c|c|}
			
			\hline 
			Field &$l_{L}$ & $l_{R}$ & $S_{L}$& $\Phi$ & $\Delta_{L}$ & $\Delta_{R}$ &  $H_{L}$ & $H_{R}$ & $\chi_{\nu}$  &$\epsilon^{'}$ & $\rho$\\ 
			\hline 
			$SU(2)_{L}$ &$2$  & $1$& $1$ & $3$ & $3$& $1$ & $2$ &$1$&$1$&$1$&$1$   \\
			\hline 
			$SU(2)_{R}$ & $1$ & $2$ &$1$& $1$ & $1$ & $3$ &  $1$ &$2$&$1$&$1$&$1$ \\
			\hline
			$U(1)_{B-L}$ &$-1$&$-1$&$0$&$2$&$2$&$2$&$-1$&$-1$&$0$&$0$&$0$  \\
			\hline 
			$A_{4}$& $3$ & $3$ & $3$ &$1$& $1$ & $1$ & $3$ &  $1$ &$3$&$1$&$1$  \\
			\hline 
			$Z_{4}$& $-1$ & $-i$ &$-i$& $1$ & $1$ & $1$ & $-i$ &$i$&$1$&$1$&$i$\\
			\hline 
		\end{tabular}
	\end{center}
	\caption{Fields and their respective transformations
		under the symmetry group of the model.} \label{tab2}
\end{table}

Now, the effective Lagrangian can be written as-
\begin{align}
	\mathcal{L}\supset
	\bar{l}_{R}(f^0_{R} \epsilon^{'} +f^\nu_{R}\chi^{\nu})\Delta_{R} l_{R}+{\bar{l}_{L}}(f^0_{L} \epsilon^{'}
	+f^\nu_{L}\chi^{\nu})\Delta_{L}l_{L}+ F\bar{l_{R}}H_{R}\rho S_{L}^{c}+h.c
\end{align} 
The role of the flavon $\epsilon^{'}$ is to break the $\mu-\tau$ symmetry of light and heavy neutrino mass matrices. In our work we take the flavon alignment to be,
$<\chi^{\nu}>\sim(1,1,1), <\rho> \sim v$.

The light neutrino mass matrix can be computed from-
\begin{equation}
	M_{L}= \nu_{L} f_{L} 
\end{equation}

Where $f_{L}$ is the Majorana Yukawa coupling. Now, the light neutrino mass matrix can be written as-

\begin{equation}
M_{L}= 
\nu_{L} f^0_{L} \begin{pmatrix}
	0  & 0 & 1 \\
	0 & 1 & 0\\
	1 & 0 & 0\\
\end{pmatrix}
+
	\nu_{L} f^{\nu}_{L} \begin{pmatrix}
		2 \chi_{1}^{\nu}  &-\chi_{3}^{\nu} &-\chi_{2}^{\nu} \\
		-\chi_{3}^{\nu} & 2\chi_{2}^{\nu} &-\chi_{1}^{\nu}\\
		-\chi_{2}^{\nu} & -\chi_{1}^{\nu} & 2\chi_{3}^{\nu}\\
	\end{pmatrix}
\end{equation}
Using the choosen flavon alignment we will get the heavy neutrino mass matrix of form given below-
\begin{equation}
M_{L} =a_{L} \begin{pmatrix}
	2z  & - z & 1 - z\\
	- z &1+ 2 z & -z\\
	1- z & -z & 2 z\\
\end{pmatrix}
\end{equation}
Where $a_{L}=\nu_{L}f^0_{L}$ and $z=\frac{f^{\nu}_{L}}{f^0_{L}}$.
Beacuse of the left-right symmetry, we can take $f_{R}=f_{L}$ and similarly we will get the Majora neutrino mass matrix of form using \eqref{eq5}
\begin{equation}
M_{R} =a_{R} \begin{pmatrix}
	2z  & - z & 1 - z\\
	- z &1+ 2 z & -z\\
	1- z & -z & 2  z\\
\end{pmatrix}
\end{equation}
Where $a_{R}=\nu_{R}f^0_{R}$ and $z=\frac{f^{\nu}_{R}}{f^0_{R}}$. So we can say that the model parameter $z=\frac{f^{\nu}_{R}}{f^0_{R}}=\frac{f^{\nu}_{L}}{f^0_{L}}$. $a_{L}$ and  $a_{R}$ are in the range of $\nu_{L}$ and $\nu_{R}$.  

Now the sterile mass matrix is
given by-
\begin{equation}
M =m_0 \begin{pmatrix}
	1  & 0 & 0\\
	0 &0  & 1\\
	0 & 1 & 0\\
\end{pmatrix}
\end{equation}
where $m_{0}$ is the free parameter of the model containing $v$ which is the vev of $\rho$. We want to work on a basis where sterile neutrino mass matrix to be diagonal. To get the diagonal sterile mass matrix we change the basis by a Unitary rotation given by-
	\begin{equation}
	U = \begin{pmatrix}
		0 & 0 & 	1  \\
		0 & 1 & 0  \\
		1 & 0 & 0\\
	\end{pmatrix}.
\end{equation}
After using this rotation one can get a diagonal sterile neutrino mass term as
At this basis  light neutrino mass matrix is given as- 
\begin{equation}
	M =m_0 \begin{pmatrix}
		1  & 0 & 0\\
		0 &1 & 0\\
		0 & 0 & 1\\
	\end{pmatrix}\label{eq9}
\end{equation}
At this basis  light neutrino mass matrix is given as-
\begin{equation}
M_{L} =a_{L} \begin{pmatrix}
	1-z & - z & 2 z\\
	- z &1+ 2 z & -z\\
	2z & -z & 1- z\\
\end{pmatrix}\label{eq7}
\end{equation}

and Majorana neutrino mass matrix-
\begin{equation}
	M_{R} =a_{R} \begin{pmatrix}
		1-z & - z & 2 z\\
		- z &1+ 2 z & -z\\
		2z & -z & 1- z\\
	\end{pmatrix}\label{eq8}
\end{equation}

Using \eqref{eq5},\eqref{eq8} and \eqref{eq9} we can have the sterile neutrino mass matrix as-\begin{equation}
	M_{S} =\frac{m_{0}^{2}}{(9z^{2}-1)a_{R}} \begin{pmatrix}
	2z+3z^2 & -z + 3 z^2 & -1 - z + 3 z^2\\
		-z + 3 z^2 &-1 + 2 z + 3 z^2 & -z + 3 z^2\\
	-1 - z + 3 z^2 & -z + 3 z^2 & 2 z + 3 z^2\\
	\end{pmatrix}\label{eq10}
\end{equation}
\section{Neutrinoless Double beta deacy in extended LRSM:}\label{sec4}
Neutrinoless double beta decay has been studied extensively within the frame work of LRSM in many literaratures \cite{Picciotto:1982qe, Ge:2015yqa,Hirsch:1996qw,Tello:2010am,BhupalDev:2013ntw,Chakrabortty:2012mh,BhupalDev:2014qbx,Bambhaniya:2015ipg}.	
In this section, we will discuss different new physics contributions to neutrinoless double beta decay process in extended LRSM scenerio. In our framework, the light neutrino mass arises from extended type-II dominant seesaw mechanism. The type-II seesaw dominance not only provides mass relation between light and heavy neutrinos but also allows large Dirac neutrino mass and thereby gives large light-heavy neutrino mixing. This light-heavy neutrino mixing plays an important role in giving sizable contributions to neutrinoless double beta decay. In this kind of scenario mass of right handed boson ($W_{R}$) is kept very high scale, so the contribution to the NDBD process due to purely right handed current is suppressed as the contribution from purely right handed current is proportional to $\frac{1}{M_{W_{R}}^{4}}$. Similarly, $\lambda$ and $\eta$ diagram gets supressed beacuse of the same argument\cite{Sruthilaya:2017vui}.

The charge current interaction for leptons and quarks which mainly governs the NDBD process are given as follows,

\begin{equation}
\mathcal{L}^q_{cc}=\frac{g_{L}}{\sqrt{2}} \bar{d}\gamma^{\mu}P_{L} W^-_{L\mu}+\frac{g_{R}}{\sqrt{2}} \bar{d}\gamma^{\mu}P_{R} W^-_{R\mu}+h.c
\end{equation}

\begin{equation}
\mathcal{L}^{lep}_{cc}=\frac{g_{L}}{\sqrt{2}}\sum_{\alpha=e,\mu,\tau} \bar{l_{\alpha}}\gamma^{\mu}P_{L} \nu_{\alpha} W^-_{L\mu}+\frac{g_{R}}{\sqrt{2}}\sum_{\alpha=e,\mu,\tau} \bar{l_{\alpha}}\gamma^{\mu}P_{R} \nu_{\alpha} W^-_{R\mu}+h.c.
\end{equation}

Mass eigenstates of neutral lepton are related to light neutrino flavor states (e,$\mu$,$\tau$) as-

\begin{equation}
\nu_{\alpha}=U_{\alpha_{i}}\nu_{i}+Z_{\alpha_{i}}N_{j}+ Y_{\alpha_{i}}S_{k}
\end{equation}
Where, $U_{\alpha_{i}}$, $Z_{\alpha_{j}}$ and $Y_{\alpha_{k}}$ are the mixing matrices of light neutrino, heavy neutrino and sterile neutrino.

The leading order contribution which contributes to the NDBD process arises from purely left handed current due to the exchange light neutrino($\nu$), heavy right handed neutrino ($N_{R}$) and sterile neutrino ($S_{L}$).The Feynmann amplitude of these contributions are proportional to-

\begin{equation}
\mathcal{A}_{\nu}^{LL}\propto G_{F}^{2} \sum_{i=1,2,3} \frac{U_{ei}^2}{p^2}
\end{equation}
\begin{equation}
	\mathcal{A}_{N}^{LL}\propto G_{F}^{2} \sum_{j=1,2,3} \big(-\frac{Z_{ej}^2}{M_{N_{j}}}\big)
\end{equation}	
\begin{equation}
	\mathcal{A}_{S}^{LL}\propto G_{F}^{2} \sum_{k=1,2,3} \big(-\frac{Y_{ek}^2}{M_{S_{k}}}\big)
\end{equation}
Where $G_{F}$ is the Fermi constant. The corresponding effective Majorana mass parameters, which are bsically the measure of lepton number violation are given as-
\begin{equation}
	|m_{ee}^{\nu}|= \sum_{i=1,2,3} U_{ei}^2 m_{\nu i},   
\end{equation}

\begin{equation}
	|m_{ee}^{N}|= <p^2>\sum_{j=1,2,3} \frac{Z_{ej}^{2}}{M_{Nj}}
\end{equation}

\begin{equation}
	|m_{ee}^{S}|= <p^2>\sum_{k=1,2,3} \frac{Y_{ek}^{2}}{M_{Sk}}
\end{equation}

where $<p^2>$  is the virtual neutrino momentum of order nuclear Fermi scale. 

\section{Light sterile neutrino in extended LRSM :\label{sec5}}		
The extended left right symmetric model discussed above can lead to a sterile neutrino of  mass at eV-GeV range depending on the parameter $m_0$. Thus it is possible to achieve a sterile neutrino in keV range within the framework which is considered as a viable dark matter candidate in present scenario. The sterile neutrino can be produced from active-sterile oscillation by Dodelson-Widrow (DW) mechanism \cite{dodelson1994sterile}. One can obtain the mass as well as the mixing of the extra sterile neutrino with the active neutrinos using \eqref{eq6}. It is evident from the expressions that both are functions of p and one can obtain the desired mass range of sterile neutrino by fine tuning $m_0$. 

There are many cosmological and astrophysical constraints on sterile neutrino dark matter \cite{Adamson:2020jvo,ng2019new,heeck2013exotic,upper}. Sterile neutrinos can be produced from Standard model plasma through their mixing with active neutrinos in the early universe. Since sterile neutrino is a fermionic dark matter candidate, lower bounds exist on its mass known as Tremaine-Gunn bound. Again, the upper limit on mass can be obtained from X-ray constraints. Direct and indirect detection of dark matter also impose significant bounds on sterile neutrino which can be seen in \cite{horiuchi2014sterile,boyarsky2009realistic,baur2017constraints,campos2016testing} .  

Any stable neutrino state with a non-vanishing mixing to the active neutrinos will be produced through active-sterile neutrino conversion. Thus the abundance is generated through the mixing between sterile and the active neutrinos. The mechanism of non resonant production of sterile neutrinos is known as Dodelson-Widrow (DW) mechanism. The resulting relic abundance can be expressed as:
\begin{equation}\label{eq:c}
\Omega_{DM}h^{2} = 1.1 \times 10^{7}\sum C_{\alpha}(m_{s})|\epsilon_{\alpha s}|^{2}{\left(\frac{m_{s}}{keV}\right)}^{2},  \alpha = e,\mu,\tau
\end{equation}
where $sin^{2}2\theta = 4 \sum|\epsilon_{\alpha s}|^{2}$ with $|\epsilon_{\alpha s}|$is the active-sterile leptonic mixing matrix element and $m_{s}$ represents the mass of the lightest sterile fermion.

The lightest sterile neutrino present in the model is not completely stable and can decay into an active neutrino through their mixing. The radiative decay of sterile neutrino  induced at one loop level  result in photon producing monochromatic X-rays. Thus there exists X-ray bounds on sterile neutrino dark matter. In LRSM, along with the bounds on bound on the mixing angle, x-ray bound leads to some constraints on the properties of the bosonic sector of the theory because of mixing of the right WR gauge bosons $W_{R}$ with the SM $W_{L}$ \cite{Bezrukov:2009th}. However, in our model, the mixing can be neglected.  The total decay width of the process $N\longrightarrow \gamma\nu$ in presence of gauge boson mixing can be written as \cite{Bezrukov:2009th},
\begin{equation}\label{eq:3}
\Gamma_{N\longrightarrow \gamma\nu} =  \frac{G_{F}^{2} \alpha m_{s}^{3}}{64 \pi^{4}}\sum\big|4 m_{l \alpha} (V_{R})_{\alpha 1}.\zeta - \frac{3}{2}\theta_{\alpha 1}m_{s}\big|^{2}
\end{equation}
Since in our model, gauge boson mixing is negligible i.e $\zeta =0$, the expression for decay width will be
\begin{equation}\label{eq:4}
\Gamma_{N\longrightarrow \gamma\nu} =  \frac{9 G_{F}^{2} \alpha m_{s}^{5}}{1024 \pi^{4}} sin^{2} 2\theta,  \alpha = e,\mu,\tau
\end{equation}

The non observation of X-ray lines from clusters provides upper limits to the active-sterile mixing angle as well as the sterile-neutrino mass. We have implemented the bounds from X-ray in our analysis. An important constraint on sterile neutrino dark matter is Ly-$\alpha$ bound. This bound provides stronger bounds on the velocity distribution of the DM particles from the effect of their free streaming on the large scale structure formation. This constraint can be converted into a bound for the mass of the DM particle which can be seen in \cite{ng2019new}. The constraint is strongly model dependent and the bounds are governed by the production mechanism of the DM candidate. In this work, We have adopted the bounds considering $XQ-100$ Ly-$\alpha$ data \cite{baur2017constraints}.

As seen from the above equations, the decay rate and as well as the relic abundance depend on mixing and mass of the DM candidate. Hence, the same set of model parameters which are supposed to produce correct neutrino phenomenology can also be used to evaluate the relic abundance and the decay rate of the sterile neutrino.

\section{Numerical results and analysis:}\label{sec6}	

In this section, we will study different new physics contribution to NDBD process in this particular framework. Here, we are interested to study different new physics contribution to the NDBD process as mentioned earlier. As the purely right handed contribution gets suppressed and dominant contribution comes from purely left handed current, we have to put Dirac neutrino mass matrix as input parameter. In our framework, we are assuming that TeV scale LRSM is originating from $SO(10)$ GUT or its sub group Pati Salam symmetry. In this limit we can take the charge lepton mass matrix and Dirac neutrino mass matrix on equal footing which is equivalent to up type quark mass matrix. The Dirac neutrino mass matrix is given as-

\begin{equation}
M_{D} =V_{CKM} M_{U} V_{CKM}^{T}\\
= \begin{pmatrix}
0.067-0.004i  & 0.302-0.022i & 0.550-0.530i \\
0.302-0.022i & 1.480 & 6.534-0.001i\\
0.550-0.530i & 6.534-0.001i & 159.72\\
\end{pmatrix} GeV
\end{equation}

The $V_{CKM}$ and up type mass matrix is given as:
\begin{equation}
V_{CKM} \\
= \begin{pmatrix}
0.97427  & 0.22534 & 0.00351-0.0033i \\
-0.2252+0.0001i & 0.97344 & 0.0412\\
0.00876-0.0032i & -0.0404-0.0007i & 0.99912\\
\end{pmatrix} GeV
\end{equation}
\begin{equation}
M_{U}= diag(2.3 MeV,1.275 GeV,173.210 GeV)
\end{equation}

The sterile and right handed neutrino mixing is coming from the model and it is a diagonal one given in \eqref{eq9}.Which is-
\begin{equation}
M =m_0 \begin{pmatrix}
1  & 0 & 0\\
0 &1 & 0\\
0 & 0 & 1\\
\end{pmatrix}
\end{equation}
The parameter $m_{o}$ is chosen to be free and we assume two different values, 50 GeV and 150 GeV  for numerical analysis. The complete diagonalization of mixing matrix give of extended LRSM with additional singlet fermion is discussed earlier and detailed analysis is given in\cite{Hati:2018tge}. Mass of light neutrino, heavy neutrino and sterile neutrino in terms of oscillation parameter can be written as:
\begin{equation}
m_\nu= M_L= U_{PMNS} m^{diag}_\nu U^T_{PMNS}
\end{equation}
\begin{equation}
M_N= M_R= \frac{\nu_{R}}{\nu_{L}}U_{PMNS} m^{diag}_\nu U^T_{PMNS}
\end{equation}
\begin{equation}
M_S= -m^{2}_{o}\frac{\nu_{L}}{\nu_{R}}U^{*}_{PMNS} (m^{diag}_\nu)^{-1} U^{\dagger}_{PMNS}
\end{equation}

$ U_{PMNS}$ is the diagonalizing matrix of the light neutrino mass matrix, $m_{\nu}$
such that,
\begin{equation}
m_{\nu}=U_{PMNS}  M_{\nu}^{(diag)}  U_{PMNS}^T
\end{equation}
where, $M_{\nu}^{(diag)} =diag(m_{1},m_{2},m_{3})$ and,
\begin{equation}
\rm U_{\text{PMNS}}=\left(\begin{array}{ccc}
c_{12}c_{13}& s_{12}c_{13}& s_{13}e^{-i\delta}\\
-s_{12}c_{23}-c_{12}s_{23}s_{13}e^{i\delta}& c_{12}c_{23}-s_{12}s_{23}s_{13}e^{i\delta} & s_{23}c_{13} \\
s_{12}s_{23}-c_{12}c_{23}s_{13}e^{i\delta} & -c_{12}s_{23}-s_{12}c_{23}s_{13}e^{i\delta}& c_{23}c_{13}
\end{array}\right) U_{\text{Maj}}
\label{matrixPMNS}
\end{equation}
is the PMNS (Pontecorvo-Maki-Nakagawa-Sakata) matrix and $c_{ij} = \cos{\theta_{ij}}, \; s_{ij} = \sin{\theta_{ij}}$ and $\delta$ is the leptonic Dirac CP phase. The diagonal matrix $\rm U_{\text{Maj}}=\text{diag}(1, e^{i\alpha}, e^{i(\beta+\delta)})$  contains the Majorana CP phases $\alpha, \beta$.

The diagonal mass matrix of the light neutrinos can be written  as, 
\begin{equation}
 M_{\nu}^{(diag)}
= \text{diag}(m_1, \sqrt{m^2_1+\Delta m_{21}^2}, \sqrt{m_1^2+\Delta m_{31}^2})
\end{equation} for normal hierarchy 
and  \begin{equation} M_{\nu}^{(diag)} = \text{diag}(\sqrt{m_3^2+\Delta m_{23}^2-\Delta m_{21}^2}, 
\sqrt{m_3^2+\Delta m_{23}^2}, m_3)\end{equation} for inverted hierarchy.

In this work, light neutrino mass directly comes from type-II dominance. Here we want to see the different contribution to the NDBD process. First of all all, we will see the case of exchange of light neutrino. The dimensionless parameter which is relevant to lepton number violating process in this process is given by

\begin{equation}
\eta_{\nu}=\frac{1}{m_e}\sum_{i=1}^{3}U^{2}_{ei}m_i=\frac{m^{\nu}_{ee}}{m_e}
\end{equation}
where $m_{e}$ is the mass of electron. And the effective mass parameter is given as
\begin{equation}
|m_{ee}^{\nu}|= \sum_{i=1,2,3} U_{ei}^2 m_{\nu i},   
\end{equation}

\begin{equation}
 m_{\nu}^{eff}=m_{1} c_{12}^2 c_{13}^2+m_{2} s_{12}^2 c_{13}^2 e^{2i\alpha} + m_{3}s_{13}^2 e^{2i\beta} 
\end{equation}
with $c_{ij} = \cos{\theta_{ij}}, \; s_{ij} = \sin{\theta_{ij}}$ are respective oscillation angle and $\alpha$ and $\beta$ are the Majorana phase.

Now, we consider the non standard contribution to the NDBD process due to purely left handed current because of the exchange of right handed neutrino. This results in dimensionless parameter which can given as-
\begin{equation}
\eta_{N}=-m_P\sum_{j=1}^{3}\frac{Z^{2}_{ej}}{M_ij}
\end{equation}
where $m_{p}$ is the mass of proton. And the effective Majorana mass parameter is-
\begin{equation}
|m_{ee}^{N}|= <p^2>\sum_{j=1,2,3} \frac{Z_{ej}^{2}}{M_{Nj}}
\end{equation}

where $<P^{2}>=\mid m_{e}m_{p}\mathcal{M}_{N}/\mathcal{M}_{\nu}\mid$ is the virtual neutrino momentum.

Now for sterile neutrino contribution, the dimensionless parameter arising from exchange of sterile neutrino can be given as-
\begin{equation}
\eta_{S}=-m_P\sum_{k=1}^{3}\frac{Y^{2}_{ek}}{M_{S_{k}}}
\end{equation}
 Effective Majorana mass parameter is -
 \begin{equation}
 |m_{ee}^{S}|= <p^2>\sum_{k=1,2,3} \frac{Y_{ek}^{2}}{M_{S_{k}}}
 \end{equation}
 
 Now the combined contribution of all the process discussed above  can be written as-
 
 \begin{equation}
 \mid m^{tot}_{ee}\mid= \sum_{i=1,2,3} U_{ei}^2 m_{\nu i}+<p^2>\sum_{j=1,2,3} \frac{Z_{ej}^{2}}{M_{Nj}}+<p^2>\sum_{k=1,2,3} \frac{Y_{ek}^{2}}{M_{S_{k}}} 
 \end{equation}

We have computed the light neutrino mass matrix from the model described in the section\ref{sec3} given in equation \eqref{eq7}. As discussed before, the structures of mass matrices of  heavy neutrino $M_{R}$ and sterile neutrino $M_{S}$ involved in analysis are constructed using the discrete flavor symmetry $A_{4}\times Z_{4}$. The light neutrino mass matrix arising from the model is consistent with non-zero $\theta_{13}$ as $A_{4}$ product rules lead to the light neutrino mass matrix in which the $\mu-\tau$ symmetry is explicitly broken.
The $3\sigma$ ranges of the mass squared differences and mixing angles from global analysis of
oscillation data are outlined as in the table\ref{tab3}.  

\begin{table}[H]
	\begin{center}
		\begin{tabular}{|c|c|c|}
			\hline 
			Oscillation parameters	& 3$\sigma$(NO) & 3$\sigma$(IO) \\ 
			\hline 
			$\frac{\Delta m_{21}^{2}}{10^{-5}eV^{2}}$	& 6.82 - 8.04 &6.82 - 8.04  \\ 
			\hline 
			$\frac{\Delta m_{31}^{2}}{10^{-3}eV^{2}}$	&   {2.431 - 2.60}  &  { 2.31-2.51}  \\ 
			\hline 
			$sin^{2}\theta_{12}$ &0.269 - 0.343  & 0.269 - 0.343 \\ 
			\hline 
			$sin^{2}\theta_{23}$ &   {0.407 - 0.612}  &   {0.411 - 0.621} \\ 
			\hline 
			$sin^{2}\theta_{13}$ &   {0.02034 - 0.02430} &   {0.02053 - 0.02436} \\ 
			\hline 
			$\frac{\delta}{\pi}$ &  {0.87 - 1.94} &   {1.12- 1.94}\\ 
			\hline 
		\end{tabular} 
	\end{center}
	\caption{Latest gobal fit neutrino oscillation  data for both mass ordering in 3-$\sigma$ range \cite{Esteban:2020cvm} }\label{tab3} 
\end{table}

\begin{figure}[H]
	\begin{center}
		\includegraphics[width=0.40\textwidth]{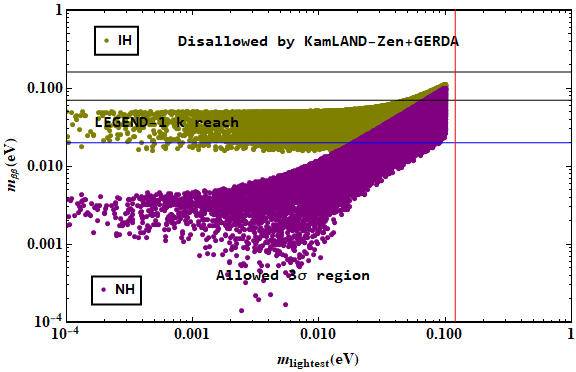}
		\includegraphics[width=0.40\textwidth]{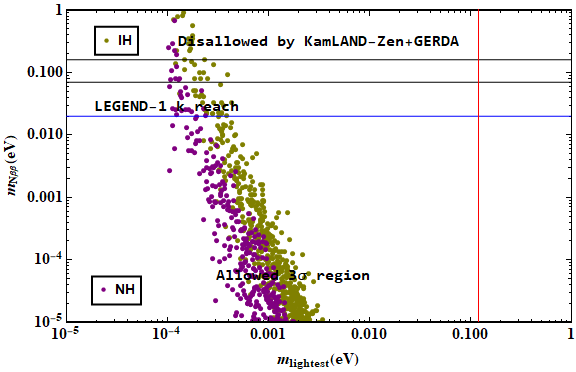}

	\end{center}	
	\begin{center}
		\caption{ Light neutrino contribution to  neutrinoless double beta decay process(left) and non standard contribution due to exchange of heavy neutrino(right) considering both NH and IH case. The
			band of black  and the blue solid horizontal line represents the KamLAND-Zen bound on the effective mass and Legend-1k reach. The red solid vertical line represents
			the Planck bound on the sum of the absolute neutrino mass respectively.}
		\label{fig20}
	\end{center}
\end{figure}

Using these values we solve for parameters of the model. From these we have evaluated the corresponding mixing matrices which are denoted as $U_{ei}$,$Z_{ej}$ and $Y_{ek}$. Then we calculated the effective mass for both the cases. During our calculation the effective mass assumes
different values depending on whether the neutrino mass states follows normal hierarchy
(NH) or inverted hierarchy (IH). The standard neutrino contribution to NDBD process as function of light neutrino mass is shown in the left side of the Fig\ref{fig20}. It is seen from the
figure that the light neutrino contribution to neutrinoless double beta decay (NDBD) can
saturate the bound imposed by KamLAND-ZEN and GERDA. We have also incorporate the future sensitivity of experiment like Legend-1k in our analysis. We have seen that for NH the contribution is well within the upper bound provided by mentioned experiments. In case of IH the contribution is well within the bound provided by KamLAND-ZEN and GERDA but some points are above the upper bound of future sensitivity of Legend-1k  reach. The right handed neutrino contribution to the NDBD process is depicted in right side of the Fig\ref{fig20}. From this figure we see that lightest neutrino mass for NH case is between the order of $10^{-2}$ to $10^{-3}$ ev and for IH can it is below $10^{-2}$ ev with effective mass ranging from $10^{-5}$ to 1 eV. For sterile contribution we have already mentioned that we will estimate this for two emperical values of $m_{o}$, 50 GeV and 150 GeV.

\begin{figure}[H]
	\begin{center}
		\includegraphics[width=0.40\textwidth]{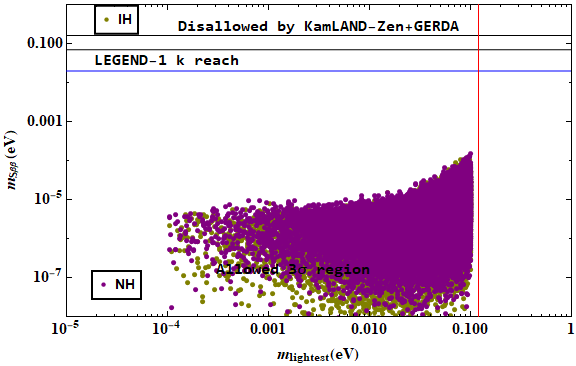}
		\includegraphics[width=0.40\textwidth]{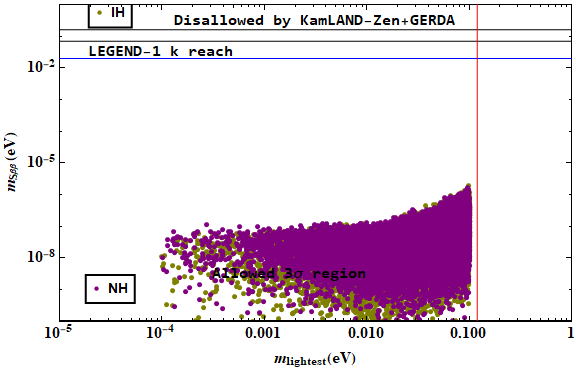}

	\end{center}	
	\begin{center}
		\caption{ Sterile neutrino contribution to  neutrinoless double beta decay process for  $m_{o}=50 $ GeV (left)  and $m_{o}=150 $ GeV(right) considering both NH and IH case. The
			band of black  and the blue solid horizontal line represents the KamLAND-Zen bound on the effective mass and Legend-1k reach. The red solid vertical line represents
			the Planck bound on the sum of the absolute neutrino mass respectively.}
		\label{fig21}
	\end{center}
\end{figure}

\begin{figure}[H]
 	\begin{center}
 		\includegraphics[width=0.40\textwidth]{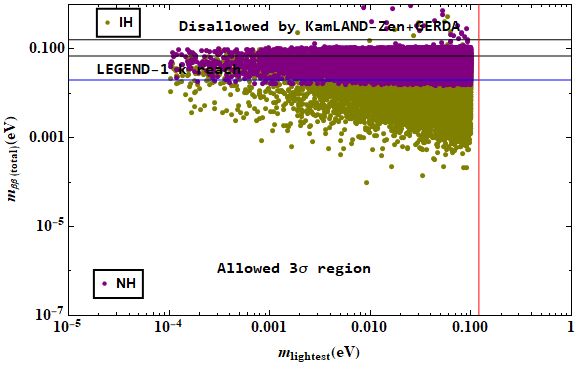}
 		\includegraphics[width=0.40\textwidth]{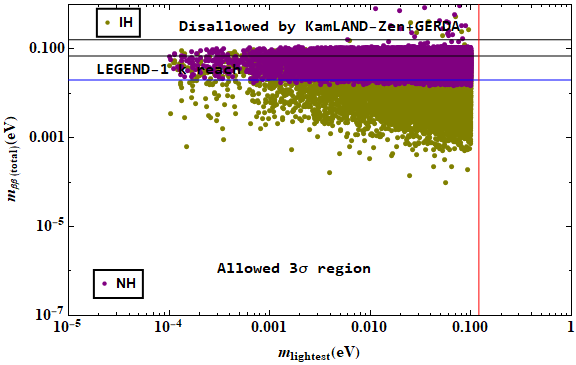}

 	\end{center}	
 	\begin{center}
 		\caption{ Total contribution to  neutrinoless double beta decay process for  $m_{o}=50 $ GeV (left)  and $m_{o}=150 $ GeV(right) considering both NH and IH case. The
 			band of black  and the blue solid horizontal line represents the KamLAND-Zen bound on the effective mass and Legend-1k reach. The red solid vertical line represents
 			the Planck bound on the sum of the absolute neutrino mass respectively.}
 		\label{fig22}
 	\end{center}
 \end{figure}

 Effective mass parameter due to the exchange of sterile neutrino considering $m_{o}=50 $ GeV as a function of lightest neutrino mass is shown in left side of the  Fig\ref{fig21}. It is seen that lightest neutrino mass ranging from $10^{-4}$ to $10^{-1}$ eV gives effective mass ranging from $10^{-4}$ to $10^{-8}$ eV for both NH and IH case which is well within the experimental limit. Again for  $m_{o}=150 $ GeV, the plot of effective mass parameter with lightest neutrino mass for sterile contribution is shown in the right side of the Fig\ref{fig21}. In this case lightest neutrino mass ranging from $10^{-4}$ to $10^{-1}$ eV gives effective mass ranging from $10^{-6}$ to $10^{-10}$ eV for both NH and IH case. So it can be inferred that on increasing the value of mass parameter of sterile neutrino the contribution from sterile neutrino to the NDBD process become smaller. The total contribution to the NDBD process considering standard, right handed neutrino and sterile neutrino for the case  $m_{o}=50 $ GeV and  $m_{o}=150 $GeV   with respect to lightest neutrino mass is shown in left and right side of Fig\ref{fig22} respectively. From these plots it can be said that there is not much difference in total contribution due to the change in the value of  $m_{o}$. For both the cases the total effective mass ranges from $10^{-4}$ to $10^{-1}$ eV for both NH and IH which is well within the bound provided by KamLAND-ZEN and GERDA. But specifically for NH in both the cases many of data points are slightly above the future sensitivity of Legend -1k reach.

\begin{figure}[H]
	\begin{center}
		\includegraphics[width=0.38\textwidth]{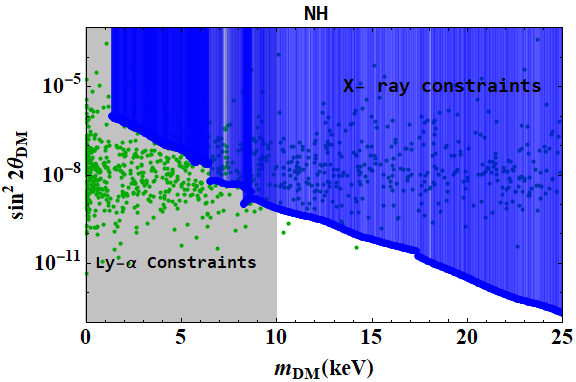}
		\includegraphics[width=0.38\textwidth]{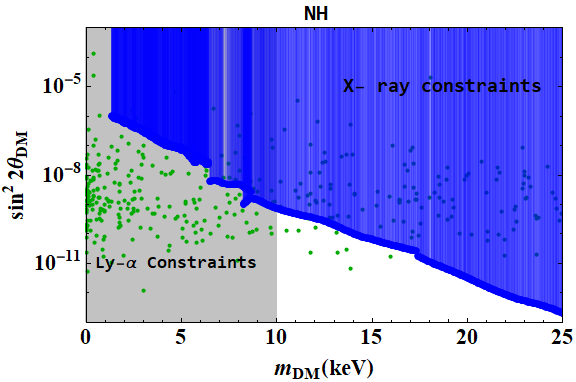}

	\end{center}	
	\begin{center}
		\caption{ Mass mixing parameter space for the DM candidate for  $m_{o}=50 $ GeV (left)  and $m_{o}=150 $ GeV(right)for NH. The cosmological bounds from Ly-$\alpha$ and X-rays have been implemented.}
		\label{fig23}
	\end{center}
\end{figure}

We also study sterile neutrino dark matter phenomenology in the framework. The same set of the evaluated model parameters are used to calculate the mass and mixing of the sterile neutrinos present in the model. We have considered the lightest sterile neutrino as potential dark matter candidate. We have obtained the relic abundance and the decay of the dark matter particle using Eq. \ref{eq:c} and Eq. \ref{eq:4} respectively. As mentioned above, we perform the dark matter study using two different values of the parameter $m_{0}$ i.e. 50 GeV and 150 GeV. The obtained results are shown from fig \ref{fig23} to fig \ref{fig27}. 

\begin{figure}[H]
	\begin{center}
		\includegraphics[width=0.40\textwidth]{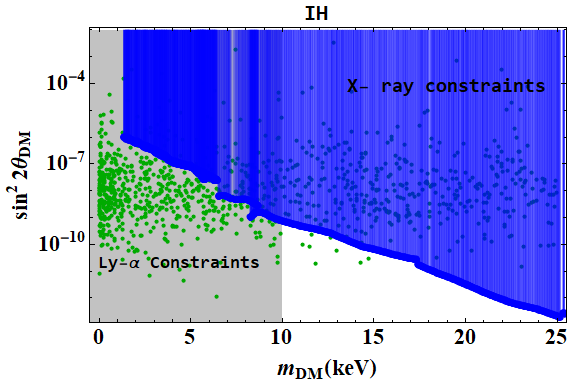}
		\includegraphics[width=0.40\textwidth]{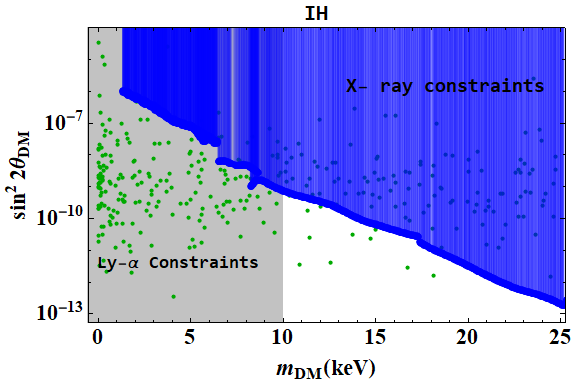}

	\end{center}	
	\begin{center}
		\caption{ Mass mixing parameter space for the DM candidate for  $m_{o}=50 $ GeV (left)  and $m_{o}=150 $ GeV(right)for IH. The cosmological bounds from Ly-$\alpha$ and X-rays have been implemented.}
		\label{fig24}
	\end{center}
\end{figure}
Fig \ref{fig23} and \ref{fig24} depict the mass mixing parameter of the sterile neutrino dark matter. After implementing the current cosmological bounds,the allowed parameter space lie within $10-25$ keV for both the cases in NH. In case of IH, the allowed parameter space is $10-24$ keV for $m_{o}= 50$ keV. It has been observed that the value of  $m_{o}$ has no significant effects on the mass and mixing limit. However, one can obtain more allowed data points for $m_{o}= 150 $ GeV as can be seen from these figures.
\begin{figure}[H]
	\begin{center}
		\includegraphics[width=0.42\textwidth]{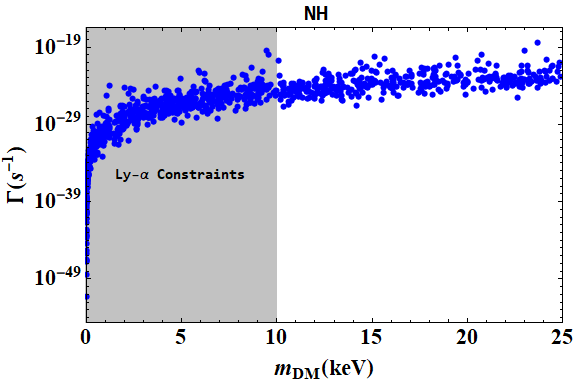}
		\includegraphics[width=0.40\textwidth]{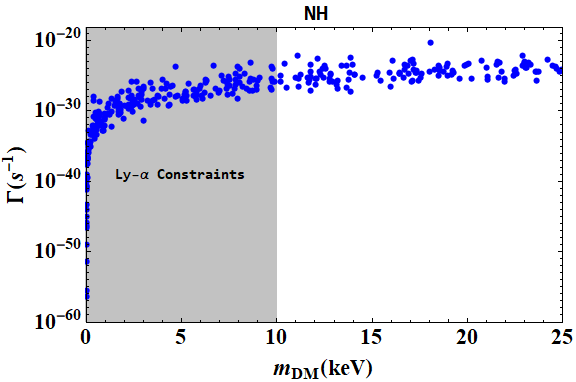}

	\end{center}	
	\begin{center}
		\caption{Decay rate of the DM candidate for  $m_{o}=50 $ GeV (left)  and $m_{o}=150 $ GeV(right)for NH. The cosmological bounds from Ly-$\alpha$ and X-rays have been implemented.}
		\label{fig25}
	\end{center}
\end{figure}

\begin{figure}[H]
	\begin{center}
		\includegraphics[width=0.40\textwidth]{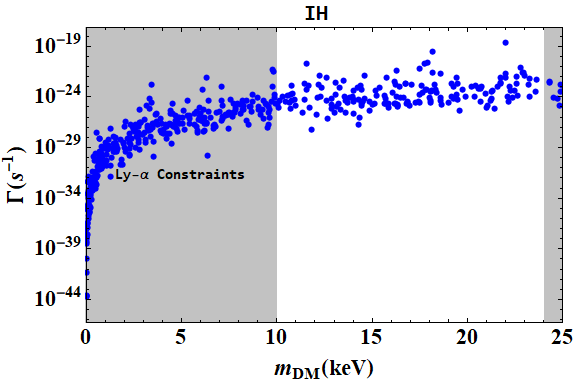}
		\includegraphics[width=0.40\textwidth]{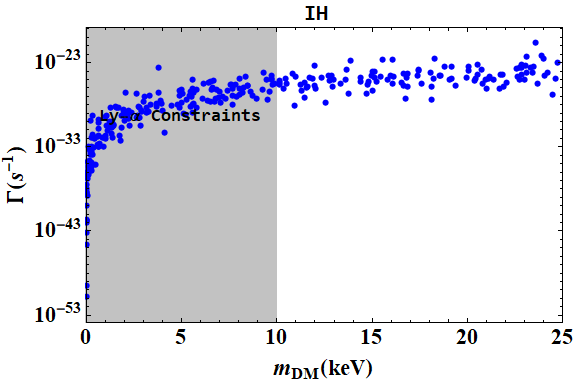}

	\end{center}	
	\begin{center}
		\caption{Decay rate of the DM candidate for $m_{o}= 50 $ GeV (left)  and $m_{o}=150 $ GeV(right)for IH. The cosmological bounds from Ly-$\alpha$ and X-rays have been implemented.}
		\label{fig25a}
	\end{center}
\end{figure}
The decay rate of the DM candidate are shown in the above figures fig \ref{fig25} and fig \ref{fig25a} for NH and IH respectively. For $m_{o}= 50 $ GeV, the decay rate for the allowed mass range is around $10^{-26}$ s$^{-1}$ and that in case of $m_{o}= 150 $ GeV is around $10^{-26}$ s$^{-1}$ in case of NH. However, the decay rates are slightly larger in case of IH for both the cases.

\begin{figure}[H]
	\begin{center}
		\includegraphics[width=0.42\textwidth]{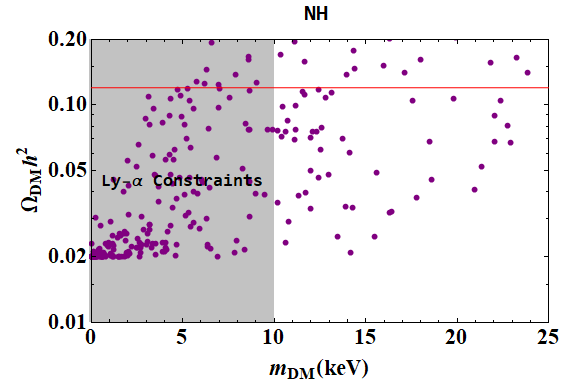}
		\includegraphics[width=0.40\textwidth]{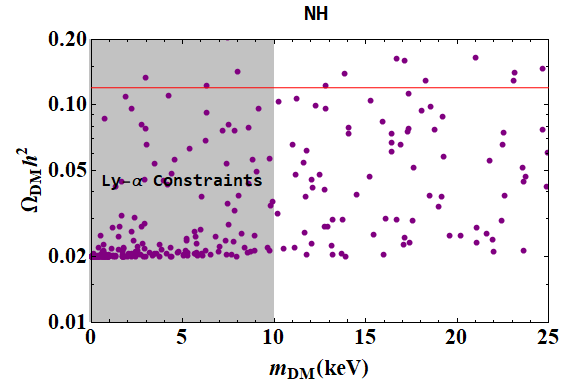}

	\end{center}	
	\begin{center}
		\caption{Sterile neutrino contribution to the total DM abundance for  $m_{o}=50 $ GeV (left)  and $m_{o}=150 $ GeV(right)for IH. The cosmological bounds from Ly-$\alpha$ and X-rays have been implemented.}
		\label{fig26}
	\end{center}
\end{figure}

\begin{figure}[H]
	\begin{center}
		\includegraphics[width=0.42\textwidth]{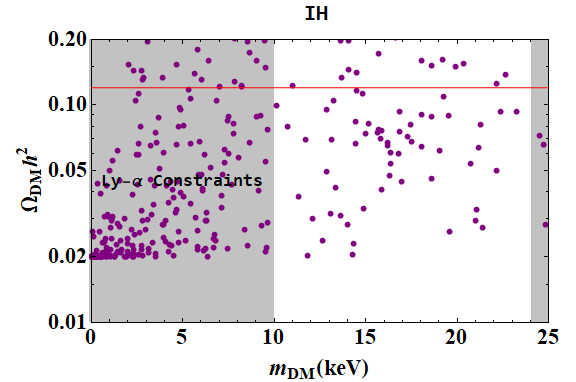}
		\includegraphics[width=0.40\textwidth]{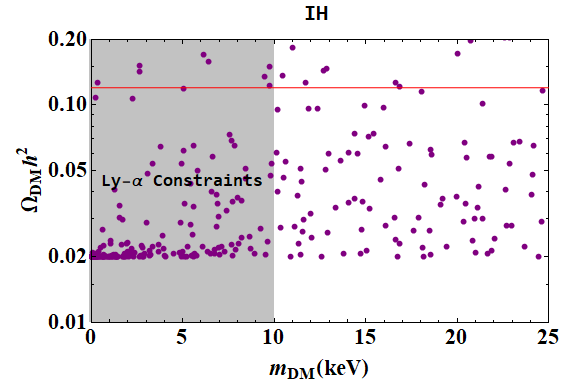}

	\end{center}	
	\begin{center}
		\caption{Sterile neutrino contribution to the total DM abundance for  $m_{o}=50 $ GeV (left)  and $m_{o}=150 $ GeV(right)for IH. The cosmological bounds from Ly-$\alpha$ and X-rays have been implemented.}
		\label{fig27}
	\end{center}
\end{figure}
Fig \ref{fig26} and fig \ref{fig27} show the relic abundance of the proposed DM candidate. The sterile neutrino DM can contribute to 17 $\%$ to almost 100 $\%$ of the total DM relic abundance in all the cases.
\begin{figure}[H]
	\begin{center}
		\includegraphics[width=0.42\textwidth]{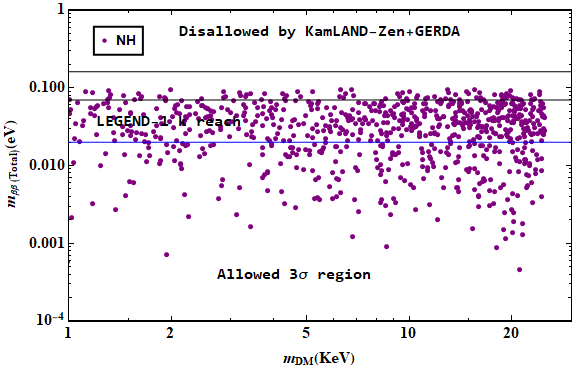}
		\includegraphics[width=0.42\textwidth]{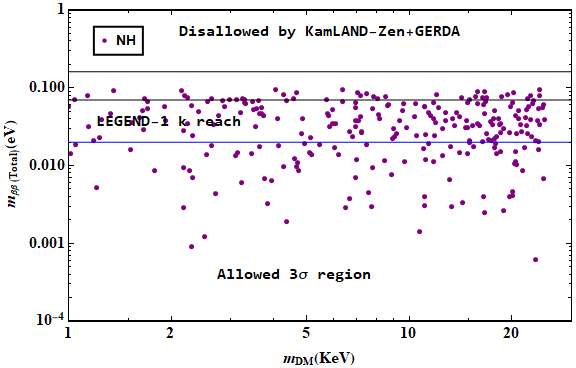}

	\end{center}	
	\begin{center}
		\caption{Correlation between DM mass and effective mass for  $m_{o}=50 $ GeV (left)  and $m_{o}=150 $ GeV(right)for NH.}
		\label{fig28}
	\end{center}
\end{figure}
\begin{figure}[H]
	\begin{center}
		\includegraphics[width=0.42\textwidth]{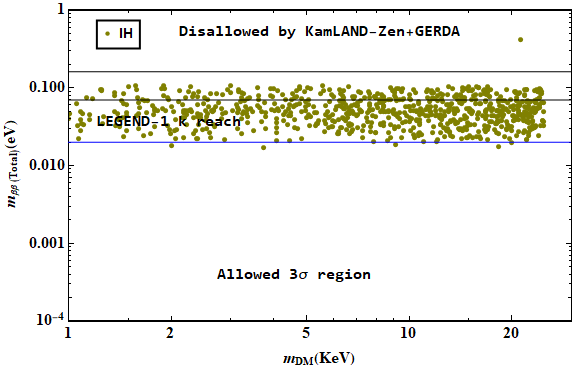}
		\includegraphics[width=0.42\textwidth]{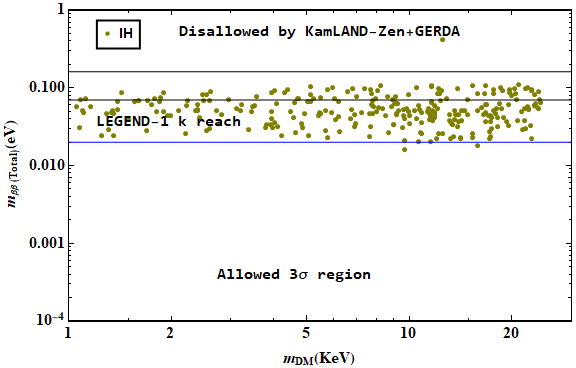}

	\end{center}	
	\begin{center}
		\caption{Correlation between DM mass and effective mass for  $m_{o}=50 $ GeV (left)  and $m_{o}=150 $ GeV(right)for IH.}
		\label{fig29}
	\end{center}
\end{figure}
Fig \ref{fig28} and fig \ref{fig29} represent the correlation of the dark matter property with the active neutrino phenomenology. It has been observed that for the allowed mass range of the DM i.e. for (10-25) keV, the effective mass range lies well within the experimental limit in all the cases. Thus the model can propose a viable DM candidate along with rich neutrino phenomenology.

\section{Conclusion}\label{sec7}	
In this work, we extend the generic left right symmetric model with an extra singlet fermion per generation. We have realized this extension with $A_{4}$ and $Z_{4}$ flavor symmetry considering type-II dominance case.
Because of the extension there will be new physics contribution to NDBD process and type-II dominance will constrain some of the contribution. We have computed all the mass matrices of light neutrino, heavy neutrino and sterile neutrino using flavor symmetry which will constrain the model. Type-II dominance gives raise to large left-right mixing which is discussed in details our work. By estimating the contribution coming from light neutrino, heavy neutrino and sterile neutrino, it is seen that all the contribution coming from these exchange are well within the experimental bound. The singlet fermion in keV range can be a viable DM candidate in the model. We have extensively studied the sterile neutrino dark matter phenomenology. It has been found that the allowed mass range lies within $10-25$ keV in this model. We have found that the DM candidate can provide significant contribution to the total relic abundance. The implications of the model on low energy processes as well as baryon asymmetry of the Universe can also be addressed which we leave for our future study. 

\bibliographystyle{utphys}
\bibliography{bb22}
\end{document}